\documentclass[prd,superscriptaddress,amsfonts,amssymb,amsmath,showpacs,twocolumn]{revtex4-2}
\usepackage{bm}
\usepackage{amsfonts}
\usepackage{latexsym}
\usepackage{graphicx}
\usepackage{amsmath}
\usepackage{palatino}
\usepackage{mathpazo}
\usepackage{textcomp}
\usepackage{float}
\usepackage{booktabs}
\usepackage{dcolumn}
\usepackage{booktabs}
\usepackage{multirow}
\usepackage{hyperref}
\hypersetup{colorlinks,citecolor=blue}
\usepackage{amsmath}
\usepackage{xcolor}
\usepackage{orcidlink}
\usepackage{epsfig}
\usepackage{caption}
\usepackage{subcaption}
\usepackage{commath}
\def\jnl@style{\it}
\def\aaref@jnl#1{{\jnl@style#1}}

\def\aaref@jnl#1{{\jnl@style#1}}

\def\aj{\aaref@jnl{AJ}}                   
\def\apj{\aaref@jnl{ApJ}}                 
\def\apjl{\aaref@jnl{ApJ}}                
\def\apjs{\aaref@jnl{ApJS}}               
\def\apss{\aaref@jnl{Ap\&SS}}             
\def\aap{\aaref@jnl{A\&A}}                
\def\aapr{\aaref@jnl{A\&A~Rev.}}          
\def\aaps{\aaref@jnl{A\&AS}}              
\def\mnras{\aaref@jnl{Mon.~Not.~Roy.~Astron.~Soc.}}             
\def\prd{\aaref@jnl{Phys.~Rev.~D}}        
\def\prc{\aaref@jnl{Phys.~Rev.~C}}  
\def\prl{\aaref@jnl{Phys.~Rev.~Lett.}}    
\def\qjras{\aaref@jnl{QJRAS}}             
\def\skytel{\aaref@jnl{S\&T}}             
\def\ssr{\aaref@jnl{Space~Sci.~Rev.}}     
\def\zap{\aaref@jnl{ZAp}}                 
\def\nat{\aaref@jnl{Nature}}              
\def\aplett{\aaref@jnl{Astrophys.~Lett.}} 
\def\apspr{\aaref@jnl{Astrophys.~Space~Phys.~Res.}} 
\def\physrep{\aaref@jnl{Phys.~Rep.}}      
\def\physscr{\aaref@jnl{Phys.~Scr}}       
\def\commat{\aaref@jnl{Comm.~Math.~Phys.}}              
\def\science{\aaref@jnl{Science}}               
\def\cqg{\aaref@jnl{Classical Quant.~Grav.}}            
\def\jpcs{\aaref@jnl{JPCS}}                                     
\def\ijmpd{\aaref@jnl{Int.~J.~Mod.~Phys.~D}}                    
\def\grg{\aaref@jnl{Gen.~Relat.~Gravit.}}               
\def\rpp{\aaref@jnl{Rep.~Prog.~Phys.}}          
\def\npa{\aaref@jnl{Nucl.~Phys.~A}}        
\def\lrr{\aaref@jnl{Living Rev.~Rel.}}                   
\def\jcap{\aaref@jnl{J.~Cosmology Astropart.~Phys.}}    
\def\rmp{\aaref@jnl{Rev.~Mod.~Phys.}}   
\def\epjc{\aaref@jnl{Eur.~Phys.~J.~C}}

\allowdisplaybreaks[1]

\begin{document}

\color{black}       

\title{Constraining the $f(R,T) = R + 2\lambda T$ cosmological model using recent observational data}

\author{N. Myrzakulov\orcidlink{0000-0001-8691-9939}}
\email[Email: ]{nmyrzakulov@gmail.com}
\affiliation{L. N. Gumilyov Eurasian National University, Astana 010008,
Kazakhstan.}
\affiliation{Ratbay Myrzakulov Eurasian International Centre for Theoretical
Physics, Astana 010009, Kazakhstan.}

\author{M. Koussour\orcidlink{0000-0002-4188-0572}}
\email[Email: ]{pr.mouhssine@gmail.com}
\affiliation{Quantum Physics and Magnetism Team, LPMC, Faculty of Science Ben
M'sik,\\
Casablanca Hassan II University,
Morocco.} 

\author{Alnadhief H. A. Alfedeel\orcidlink{0000-0002-8036-268X}}%
\email[Email:]{aaalnadhief@imamu.edu.sa}
\affiliation{Department of Mathematics and Statistics, Imam Mohammad Ibn Saud Islamic University (IMSIU),\\
Riyadh 13318, Saudi Arabia.}
\affiliation{Department of Physics, Faculty of Science, University of Khartoum, P.O. Box 321, Khartoum 11115, Sudan.}
\affiliation{Centre for Space Research, North-West University, Potchefstroom 2520, South Africa.}

\author{E. I. Hassan\orcidlink{0000-0000-0000-0000}}%
\email[Email:]{eiabdalla@imamu.edu.sa}
\affiliation{Department of Mathematics and Statistics, Imam Mohammad Ibn Saud Islamic University (IMSIU),\\
Riyadh 13318, Saudi Arabia.}

\date{\today}

\begin{abstract}
In this paper, we consider a comprehensive investigation of the cosmological model described by $f(R,T) = R + 2\lambda T$ (where $\lambda$ represents a free parameter) in light of the most recent observational data. By constraining the model using $Hubble$ and $Pantheon$ datasets, we determine its compatibility with the observed behavior of the Universe. For this purpose, we adopt a parametric form for the effective equation of state (EoS) parameter. This parametric form allows us to describe the evolution of the EoS parameter with respect to redshift and investigate its behavior during different cosmic epochs. The analysis of the deceleration parameter reveals an accelerating Universe with a present value of $q_0=-0.64^{+0.03}_{-0.03}$, indicating the current phase of accelerated expansion. The transition redshift is found to be $z_{tr}=0.53^{+0.04}_{-0.03}$, marking the epoch of transition from deceleration to acceleration. We also analyze the evolution of important cosmological parameters including density parameter, pressure, effective EoS, and stability. These findings collectively demonstrate the viability of the $f(R,T)$ cosmological model as a robust candidate capable of engendering the requisite negative pressure, thereby efficiently propelling cosmic expansion. Moreover, the undertaken stability analysis underscores the model's stability within the broader cosmic landscape. 
By providing the best-fit values for the coupling parameter $\lambda$, this approach motivates and encourages further explorations into the extensive landscape of this model and its potential applications across diverse realms of cosmology and astronomy.

\end{abstract}

\maketitle

\section{Introduction}
\label{sec1}

Astronomical observations from various sources such as Type Ia Supernovae (SNeIa) \cite{Perlmutter/1999,Riess/1998,Riess/2004}, Cosmic Microwave Background Radiation (CMBR) \cite{Komatsu/2011,Huang/2006}, and Large-Scale Structures (LSS) \cite{Koivisto/2006, Daniel/2008} have provided evidence for the transition of the Universe from an early deceleration phase to a recent acceleration phase. This transition has sparked the search for the underlying cause of the late-time cosmic accelerated expansion, which remains a significant challenge in modern cosmology. The dominant component driving this expansion is referred to as Dark Energy (DE), an elusive form of energy that remains poorly understood \cite{Peebles/1988,Ratra/1988,Steinhardt/1999}. While the inclusion of DE, such as the cosmological constant, has been successful in explaining the accelerated expansion of the Universe, it is plagued by certain theoretical challenges. Two prominent issues are the problems of cosmic coincidence and fine-tuning. The cosmic coincidence problem, questions the remarkable coincidence that the energy densities of DE and matter are of the same order of magnitude at the present epoch, despite evolving differently over cosmic time. On the other hand, the fine-tuning problem refers to the difficulty in explaining why the observed value of DE density is so incredibly small compared to theoretical predictions. \cite{Peebles/2003,Sahni/2000}. 

An alternative avenue to address the challenges posed by DE is to explore modifications to the gravitational sector of the Einstein-Hilbert (EH) action. This approach, known as Modified Theories of Gravity (MTGs), offers a different perspective on the nature of cosmic acceleration. Geometrically MTGs offer an extended framework beyond General Relativity (GR), where the EH action can be modified by replacing the Ricci scalar $R$ (or curvature scalar) with a more general function. These modifications can involve coupling matter with geometry through various scalar quantities. Several examples of such MTGs include $f(R)$ gravity, $f(\mathcal{T})$ gravity, $f(Q)$ gravity, and $f(R,T)$ gravity. In $f(R)$ gravity, the Ricci scalar $R$ is replaced with a more general function, allowing for deviations from standard GR predictions \cite{carroll04, nojiri07, bertolami07,Dixit}. Similarly, in $f(\mathcal{T})$ gravity, the torsion scalar $\mathcal{T}$ is involved in the modification of the gravitational action \cite{bengocheu09,linder10,Chen,Anagnostopoulos}. $f(Q)$ gravity incorporates the non-metricity scalar $Q$ in the gravitational action, introducing additional geometric terms \cite{Jimenez1,Jimenez2,Banerjee,Pradhan6}. Among these geometrically MTGs, $f(R,T)$ theory of gravity (where $R$ is the Ricci scalar and $T$ is the trace of the energy-momentum tensor) \cite{Harko2011} has gained significant attention from cosmologists and astrophysicists due to its ability to address various cosmological and astrophysical issues.

The versatility of $f(R,T)$ gravity has made it a subject of extensive investigation, with researchers exploring its implications for different phenomena in cosmology and astrophysics. This modified theory offers a promising avenue to tackle some of the challenges and unanswered questions in these fields. As a result, $f(R,T)$ gravity has garnered considerable interest and has been the focus of numerous studies in recent years \cite{Sahu17,Mishra16,Moraes17,Yousaf17,Pradhan5,Pretel}. Recently, da Silva et al. \cite{Silva} explored the behavior of rapidly rotating neutron stars within the framework of $f (R, T)$ gravity. The study investigates the effects of $f(R,T) = R + 2\lambda T$ gravity on the structure and properties of neutron stars, which are extremely dense and compact astrophysical objects. Vinutha et al. \cite{Vinutha} conducted a study where they analyzed the field equations and derive the dynamical equations for anisotropic perfect fluid cosmological models in $f (R, T)$ gravity. They investigate the solutions and explore the implications of anisotropy on the evolution of the scale factor, energy density, and other cosmological parameters. Bishi et al. \cite{Bishi} investigated the existence of the Gödel Universe within different functional forms of $f(R, T)$ gravity. Their study focused on exploring the possibility of constructing cosmological solutions that resemble the Gödel Universe, which is characterized by the presence of rotation and closed timelike curves.

Despite numerous observations confirming the existence of DE, its underlying nature still eludes us. The criterion for the accelerated expansion of the Universe is determined by the Equation of State (EoS) parameter, where $\omega<-\frac{1}{3}$. Understanding the gravitational dynamics of the Universe necessitates an exploration of the physics governing DE and its corresponding EoS \cite{Carroll/2003,Gong/2007,Huang/2008}. Therefore, this study aims to combine the parametrized EoS with the modified $f(R,T)$ gravity, bridging the gap between these two aspects to gain deeper insights into the behavior of the Universe. In the study of DE, various parametrizations of the EoS have been proposed to capture its evolving nature. One widely used parametrization is the Chevallier-Polarski-Linder (CPL) parametrization: $\omega(z)=\omega_{0}+\omega_{1}\frac{z}{1+z}$, which is based on a simple Taylor expansion of the EoS in terms of the scale factor \cite{CPL1,CPL2}. While the CPL parametrization proves to be a reliable choice for describing the behavior of the Universe at early $(z\to \infty)$ and present $(z=0)$ epochs, it exhibits a divergence at future times. Specifically, at a redshift of $z = -1$, the CPL parametrization encounters issues. However, it should be indicated that the CPL parametrization performs effectively at large redshifts and serves as a suitable approximation for slow-roll DE scalar field models. In addition to the CPL parametrization, several other parametrizations have also been proposed, such as the Jassal-Bagla-Padmanabhan (JBP) parametrization i.e. $\omega(z)=\omega_{0}+\omega_{1}\frac{z}{(1+z)^2}$ \cite{JBP}, and the Ma-Zhang (MZ) parametrization, which offers a unique approach by utilizing a logarithmic and oscillating form to describe the behavior of the EoS, i.e. $\omega(z)=\omega_{0}+\omega_{1}(\frac{\ln(2+z)}{1+z}-\ln2)$ and $\omega(z)=\omega_{0}+\omega_{1}(\frac{\sin(1+z)}{1+z}-\sin(2))$, respectively \cite{MZ}.

Motivated by the previous discussion and the need to understand the behavior of the DE, we delve into a comprehensive investigation of the cosmological model characterized by $f(R,T) = R + 2\lambda T$, where $\lambda$ represents a free parameter. Our analysis is guided by the desire to explore the compatibility of this model with the latest observational data. By incorporating the $Hubble$ and $Pantheon$ datasets, we aim to shed light on the observed evolution of the Universe and assess the viability of the proposed model. Furthermore, we employ a parametric form for the effective EoS parameter, which allows us to examine the dynamics of the model and compare its predictions with the empirical findings from observational data.

This paper is divided into the following sections: Sec. \ref{sec2} presents a detailed discussion on the formalism of $f(R, T)$ gravity. In Sec. \ref{sec3}, we derive the expression for the Hubble parameter within the Friedmann-Lemaitre-Robertson-Walker (FLRW) framework, utilizing a one-parameter EoS. The process of constraining the model parameters using different datasets, including the $Hubble$ data, the $Pantheon$ data, and the joint $Hubble+Pantheon$ datasets through the employment of the MCMC technique, is described in Sec. \ref{sec4}. The behavior of various cosmological parameters is analyzed in Sec. \ref{sec5}. We also discuss a stability analysis of the model in Sec. \ref{sec6}. Finally, Sec. \ref{sec7} provides a comprehensive discussion of the obtained results.

\section{A Comprehensive Overview of $f(R,T)$ Theory and its Fundamental Aspects}
\label{sec2}

In $f(R, T)$ gravity theory, the EH action is augmented with geometric modifications, resulting in a revised framework for describing the gravitational dynamics of the Universe. By introducing additional terms that depend on the Ricci scalar $R$ and the trace of the energy-momentum tensor $T$, the action is given by
\begin{equation}
\mathbb{S}=\frac{1}{2\kappa}\int  f(R,T)\sqrt{-g}d^{4}x +\int \mathcal{L}_{m}\sqrt{-g}d^{4}x.\label{action}
\end{equation}

Furthermore, by varying the metric tensor $g_{\mu\nu}$, we can derive the gravitational field equation for $f(R, T)$ gravity from the modified action as,
\begin{multline}
f_{R}(R,T)R_{\mu\nu}-\frac{1}{2}f(R,T)g_{\mu\nu}+(g_{\mu\nu}\Box -\nabla _{\mu}\nabla
_{\nu})f_{R}(R,T) \\ = \kappa T_{\mu\nu}-f_{T}(R,T)T_{\mu\nu}- f_{T}(R,T)\Theta _{\mu\nu}.
\end{multline}

Here, we define several important quantities. First, let's denote the partial derivative of $f(R,T)$ with respect to $R$ as $f_{R}(R,T)=\frac{\partial f(R,T)}{\partial R}$ and the  partial derivative with respect to $T$ as $f_{T}(R,T)=\frac{%
\partial f(R,T)}{\partial T}$. The symbol $\Box$ represents the d'Alembertian operator $\Box \equiv \nabla ^{\mu}\nabla _{\mu}$, where $\nabla _{\mu}$ represents the covariant derivative. The constants $\kappa$ are defined as $\kappa=\frac{8\pi G}{c^4}$, where $G$ is the Newtonian Gravitational constant and $c$ is the speed of light in a vacuum.

For the energy-momentum tensor of a perfect fluid distribution in the Universe, we have $T_{\mu\nu}=-pg_{\mu\nu}+(\rho+p)u_\mu u_\nu$. Here, $\rho$ represents the energy density, $p$ represents the pressure, and $u^\mu$ is the 4-velocity of the fluid, satisfying the condition $u_\mu u^\nu=1$ in comoving coordinates.

In addition, we introduce the tensor $\Theta_{\mu\nu} =g^{\alpha \beta} \frac{\delta T_{\alpha \beta}}{\delta g^{\mu\nu}}$, which is derived from the matter Lagrangian $\mathcal{L}_m$. Following Harko et al. \cite{Harko2011}, we choose the matter Lagrangian as $\mathcal{L}_m=-p$, resulting in $\Theta_{\mu\nu}=-pg_{\mu\nu}-2T_{\mu\nu}$.

Furthermore, it is noteworthy to highlight that the covariant divergence of the matter-energy-momentum tensor within the framework of the $f(R,T)$ theory can be represented as,
\begin{equation}
    \nabla ^{\mu }T_{\mu \nu }=-\frac{\kappa }{1+\kappa f_{T} }%
\left[ T_{\mu \nu }\nabla ^{\mu }f_{T} +g_{\mu \nu }\nabla
^{\mu }\left( f_{T} p\right) \right]. 
\end{equation}

Therefore, the equation presented above serves as a clear illustration of a fundamental aspect within the framework of the $f(R,T)$ gravity theory. Specifically, it points to a noteworthy departure from the conventional conservation behavior of the matter-energy-momentum tensor. In more conventional scenarios, such as in GR, the covariant divergence of the matter-energy-momentum tensor typically vanishes, implying a strict conservation law. However, in the context of the $f(R,T)$ gravity theory, this familiar conservation property is no longer maintained. The equation $\nabla ^{\mu }T_{\mu \nu }\neq 0$ distinctly signifies that the matter-energy-momentum tensor does not adhere to the expected conservation behavior. This lack of conservation can be interpreted as the presence of an additional force acting on massive test particles, leading to non-geodesic motion. From a physical standpoint, it signifies the flow of energy into or out of a specified volume within a physical system. Additionally, the non-zero right-hand side of the energy-momentum tensor signifies the occurrence of transfer processes or particle production within the system. Notably, in the absence of $f_T$ terms in the equation, the energy-momentum tensor becomes conserved \cite{Harko2011}.

By considering these definitions and relationships, we can further analyze and understand the properties and behavior of the $f(R,T)$ gravity theory. 
The functional $f(R,T)$ provides flexibility in choosing various viable models within the $f(R,T)$ gravity framework. In our current study, we specifically consider the functional $f(R,T)=R+2f(T)$, where $f(T)$ represents an arbitrary function of the trace of the energy-momentum tensor. We explore the implications and consequences of $f(R,T)$ gravity by employing this particular form. With this choice, the corresponding field equations can be derived as,
\begin{equation}
R_{\mu\nu}-\frac{1}{2}Rg_{\mu\nu}=\kappa T_{\mu\nu}+2f_T T_{\mu\nu}+\left[f(T)+2pf_T\right]g_{\mu\nu}. \label{FE}
\end{equation}

\section{The cosmological model}
\label{sec3}

In this work, we assume the specific functional form $f(T)=\lambda T$, where $\lambda$ is a constant. This specific functional expression for $f(R, T)$ has received significant attention within the existing literature \cite{Harko2011,Sahu17,Mishra16}. Its widespread study enhances the comparability of our findings with those obtained by other researchers working within this framework. For instance, investigations into cosmic acceleration, as demonstrated in \cite{Pradhan5}, can be readily linked to our results. Additionally, this choice holds the notable advantage of avoiding the introduction of higher-order derivatives into the field equations. However, it is important to acknowledge that our choice is not exhaustive, and indeed, there are various alternative functional forms for $f(R, T)$ that could be explored \cite{Moraes17,Yousaf17,Vinutha}.

In addition, we consider the implications of this choice on the field equations for a flat homogeneous and isotropic FLRW metric,
\begin{equation}
ds^2=dt^2-a^2(t)\left[dx^2+dy^2+dz^2\right] \label{FLRW}
\end{equation} 
where $a(t)$ represents the scale factor of the Universe. By plugging the FLRW metric \eqref{FLRW} into the field equations of $f(R,T)$ gravity, we can derive the specific form of the field equations for this choice of $f(T)$ as,
\begin{equation}
3H^2=(1+3\lambda)\rho-\lambda p, \label{F1}
\end{equation}
\begin{equation}
2\dot{H}+3H^2=\lambda \rho-(1+3\lambda)p.\label{F2}
\end{equation}
where $H=H(t)=\frac{\dot{a}}{a}$ represents the Hubble parameter, which characterizes the rate of expansion of the Universe. In our analysis, we adopt a unit system where we set $\kappa=1$. 

From Eqs. (\ref{F1}) and (\ref{F2}), the energy density and pressure can be determined as,
\begin{equation}
\rho =\frac{(3+6\lambda)H^2-2\lambda \dot{H}}{(1+3\lambda)^2-\lambda^2},\label{rho} 
\end{equation}

\begin{equation}
p = \frac{-(3+6\lambda)H^2-2(1+3\lambda) \dot{H}}{(1+3\lambda)^2-\lambda^2}.\label{p}
\end{equation}

The effective equation of state (EoS) parameter, which represents the ratio of pressure to energy density for all the cosmological components, including DE, matter, and radiation i.e. $\omega_{eff}=\frac{p}{\rho}$, can be expressed as:
\begin{equation}
    \omega_{eff}=\frac{-(3+6\lambda)H^2-2(1+3\lambda) \dot{H}}{(3+6\lambda)H^2-2\lambda \dot{H}}.
    \label{eff1}
\end{equation}

To facilitate the comparison between theoretical predictions and cosmological observations, we introduce the redshift variable $z$ as an independent variable instead of the conventional time variable $t$. The redshift is defined as
\begin{equation}
    1+z=\frac{1}{a(t)}.
\end{equation}

By normalizing the scale factor such that its present-day value is one ($a(0) = 1$), we can establish a relationship between the derivatives with respect to time and the derivatives with respect to the redshift. Thus, the time derivative of the Hubble parameter can be expressed in the following form:
\begin{equation}
\label{Hdz}
\dot{H}=\frac{dH}{dt}=-\left(1+z\right)H(z)\frac{dH}{dz}.
\end{equation}

In order to obtain the solution for the Hubble parameter, an additional ansatz is required. In this study, we adopt a specific parametrization for the effective EoS. We consider a parametric form for the effective EoS parameter $\omega_{eff}$ in terms of the redshift $z$, incorporating a single parameter, expressed as \cite{Lapola}:
\begin{equation}
    \omega_{eff}=\frac{1}{3} \left[1-\frac{4}{1+\chi  (1+z)^4}\right]
    \label{eff2}
\end{equation}
where $\chi$ is a constant. The choice of the specific parametric form emerges from a synthesis of theoretical considerations and empirical insights. This form aligns with the standard cosmological model and adeptly captures the evolution of the EoS parameter across varying cosmic epochs \cite{Tanabashi}. The selection of this form allows us to depict the cosmic landscape as it transitions from DE domination to matter and radiation domination, providing an inclusive framework for comprehensive analysis. For the present redshift (i.e. at $z=0$), the effective EoS parameter is expected to be less than $-1/3$, indicating the era dominated by DE. The exact value of $\omega_{eff}$ is $\omega_{eff}(z=0)=\frac{1}{3} \left[1-\frac{4}{1+\chi}\right]$
i.e. depends on the value of $\chi $. In the past, the effective EoS parameter approaches zero i.e. $\omega_{eff}(z >0)=0$, which is consistent with the matter-dominated phase. For larger redshift values, the EoS parameter converges to $\omega_{eff}(z \to \infty)=1/3$, representing the era dominated by radiation energy. This parametric form thus captures the expected behavior of the EoS parameter in different cosmic epochs, aligning with the predictions of the standard cosmological model. Importantly, this choice stands on the shoulders of related parametrization schemes explored in the literature \cite{Mussatayeva}. These schemes encompass a diverse array of theoretical and observational perspectives, enhancing our understanding of the dynamics driving cosmic evolution. Mukherjee's study concentrated on the acceleration of the universe and offered a reconstruction of the effective EoS \cite{Mukherjee}. Moreover, the same form has been employed across diverse theories of modified gravity \cite{Koussour1,Arora}.

By combining Eqs. (\ref{eff1}) and (\ref{eff2}), we derive the following differential equation:
\begin{equation}
    \frac{6 \left(H^2+\dot{H}\right) \lambda +3 H^2+2 \dot{H}}{2 \dot{H} \lambda -3 H^2 (1+2 \lambda)}+\frac{4}{3+3 \chi  (1+z)^4}-\frac{1}{3}=0.
    \label{diff}
\end{equation}

Using Eq. (\ref{Hdz}) and solving Eq. (\ref{diff}) leads to the solution:
\begin{equation}
    H(z)=H_0 \left(\frac{12 \lambda +(3+8 \lambda) \chi  (1+z)^4+3}{8 \lambda  \chi +12 \lambda +3 \chi +3}\right)^{l}
\end{equation}
where $l=\frac{3+6 \lambda}{6+16 \lambda }$ and $H_0=H(z=0)$ represent the present value of the Hubble parameter.

The deceleration parameter $q$, which is defined as $q=-1-\frac{\dot{H}}{H^2}$, can be obtained by the following expression:
\begin{equation}
    q(z)=-1+\frac{6 (1+2 \lambda) \chi  (1+z)^4}{12 \lambda +(3+8 \lambda) \chi  (1+z)^4+3}.
\end{equation}

\section{Analysis of Observational Data}
\label{sec4}

In this section, we present an overview of the cosmological data utilized in our investigation. To constrain the parameters in the $H(z)$ model, we employ a range of contemporary observational data and utilize the MCMC (Markov Chain Monte Carlo) technique. Our focus is on data that provides insights into the expansion history of the Universe, particularly those pertaining to the distance-redshift relation. Specifically, we incorporate data from early-type galaxies, which provide direct measurements of the Hubble parameter $H(z)$ \cite{Moresco/2015,Moresco/2018}. In addition, we incorporate data from SNeIa using the $Pantheon$ samples, which encompass observations from the Supernova Legacy Survey (SNLS), Sloan Digital Sky Survey (SDSS), Hubble Space Telescope (HST) survey, and the Panoramic Survey Telescope and Rapid Response System (Pan-STARRS1) \cite{Scolnic/2018,Chang/2019}. By leveraging this diverse range of observational data, we aim to obtain robust constraints on the parameters of the $H(z)$ model and gain further insights into the expansion history of our Universe.

\subsection{Cosmic Chronometers: Estimating the Hubble parameter from early-type galaxies}

To estimate the Hubble parameter for early-type galaxies with passive evolution, we rely on the prediction of their differential evolution. The compilation of such data is commonly referred to as cosmic chronometers (CC) \cite{Moresco/2015,Moresco/2018}. In this study, we utilize a sample of cosmic chronometers covering the redshift range of $0<z<1.97$. To assess the constraints on the model parameters, we employ the chi-squared ($\Tilde{\chi}^{2}$) estimator:
\begin{equation}
\Tilde{\chi}^{2}_{Hubble} = \sum_{i=1}^{31} \frac{\left[H(\theta_{s}, z_{i})-
H_{obs}(z_{i})\right]^2}{\sigma(z_{i})^2}.
\end{equation}

Here, $H(\theta_{s},z_{i})$ represents the theoretical prediction of the Hubble parameter at redshift $z_{i}$, while $H_{obs}(z_{i})$ represents the observed values. The term $\sigma^{2}_{Hub}(z_{i})$ denotes the standard error associated with the measured values of $H_{obs}(z_{i})$ and $\theta_{s}=(H_0,\lambda,\chi)$ is the parameter space of the cosmological model.

\subsection{The $Pantheon$ compilation: Constraints from SNeIa}

The $Pantheon$ compilation, as presented by Scolnic et al. \cite{Scolnic/2018}, is a comprehensive and up-to-date collection of SNeIa data. In our analysis, we utilize a dataset consisting of 1048 SNeIa spanning a redshift range of $0.01<z<2.26$. To quantify the agreement between the theoretical predictions and the observed SNeIa data, we employ the $\Tilde{\chi}^{2}$ statistic:
\begin{equation}
\Tilde{\chi}^{2}_{Pantheon} =\sum_{i,j=1} ^{1048} \Delta \mu_{i} \left(
C_{Pantheon}^{-1}\right)_{ij} \Delta \mu_{j}.
\end{equation}

Here, $\Delta \mu_{i}=\mu_{\mathrm{th}}-\mu_{\mathrm{obs}}$ represents the difference between the theoretical and observed distance modulus, where $\mu = m_{B}-M_{B}$ represents the difference between the apparent magnitude $m_{B}$ and the absolute magnitude $M_{B}$. The term $C_{Pantheon}^{-1}$ corresponds to the inverse of the covariance matrix of the $Pantheon$ sample. In addition, the theoretical value of the distance modulus is computed using the formula:
\begin{equation}
\mu _{th}(z)=5log_{10}\frac{d_{L}(z)}{1Mpc}+25,
\end{equation}%
where $d_{L}(z)$ denotes the luminosity distance that incorporates the attenuation of light due to the expansion of the Universe. The luminosity distance is evaluated by integrating the expression:
\begin{equation}
d_{L}(z,\theta_{s})=(1+z)\int_{0}^{z}\frac{dy}{H(y,\theta_{s})},
\end{equation}%

\subsection{Joint constraints and likelihood functions}

To obtain combined constraints for the parameters $\theta_{s}=(H_0,\lambda,\chi)$ from the $Hubble$ and $Pantheon$ samples, we use the total likelihood function. The relevant likelihood and $\chi^2$ functions are defined as follows:
\begin{eqnarray}
\mathcal{L}_{joint} &=& \mathcal{L}_{Hubble} \times \mathcal{L}_{Pantheon},\\
\Tilde{\chi}^{2}_{joint} &=& \Tilde{\chi}^{2}_{Hubble} + \Tilde{\chi}^{2}_{Pantheon}.
\end{eqnarray}

Here, $\mathcal{L}_{Hubble}$ and $\mathcal{L}_{Pantheon}$ represent the likelihood functions for the $Hubble$ and $Pantheon$ samples, respectively. The total likelihood function $\mathcal{L}_{joint}$ is obtained by taking the product of these individual likelihood functions. Similarly, the $\Tilde{\chi}^{2}_{joint}$ is obtained by summing the individual $\Tilde{\chi}$ values for the $Hubble$ and $Pantheon$ samples.

To determine the constraints on the model parameters, we minimize the corresponding $\Tilde{\chi}$ function using the MCMC method and the $emcee$ library \cite{Mackey/2013}. The MCMC technique allows us to explore the parameter space and obtain a statistical distribution of parameter values consistent with the observational data. The results of this analysis can be found in Tab. \ref{tab}. Further, Fig. \ref{Hubble} and Fig. \ref{Mu} illustrate the error bar fits for the considered model, as well as the $\Lambda$CDM model with specific parameter values $\Omega_{m_0}=0.315$, and $H_0=67.4$ $km/s/Mpc$ \cite{Planck2020}. These figures provide visual comparisons between the model predictions and the observational data. 

Furthermore, Fig. \ref{Combine} displays the $1-\sigma$ and $2-\sigma$ contour plots for the $Hubble$, $Pantheon$, and joint observational data, demonstrating the range of parameter values consistent with the observations. An observable disparity in the values of the Hubble constant $H_0$ becomes evident when contrasting across different datasets. This discrepancy arises from the distinct nature of the two datasets and the methodologies employed for their analysis. While CC directly measure the Hubble parameter at different redshifts, providing a robust and model-independent determination of $H_0$, SNeIa data involve the intricate cosmic distance ladder and necessitate modeling assumptions.

\begin{widetext}

\begin{figure}[H]
\centering
\includegraphics[scale=0.60]{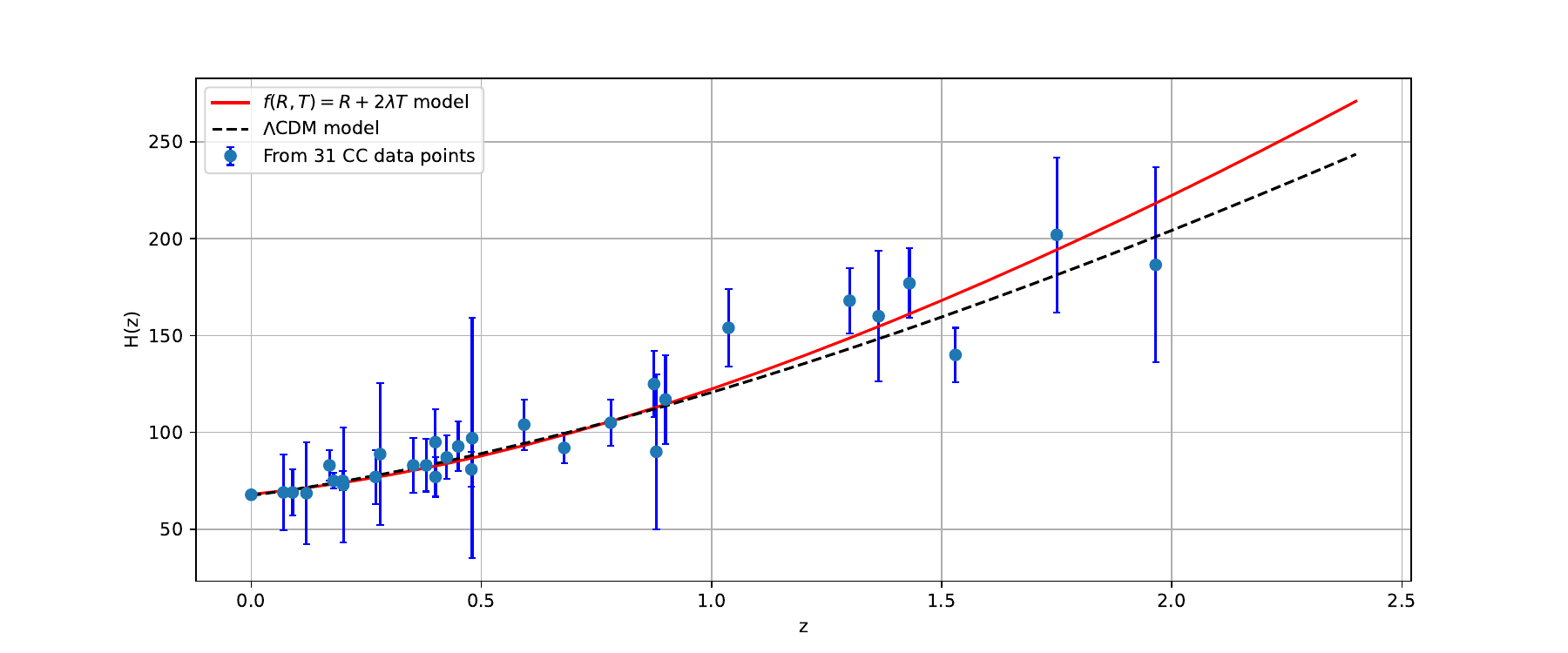}
\caption{Evolution of the Hubble parameter $H(z)$ with redshift $z$ in comparison to $\Lambda$CDM model.}
\label{Hubble}
\end{figure}	

\begin{figure}[H]
\centering
\includegraphics[scale=0.60]{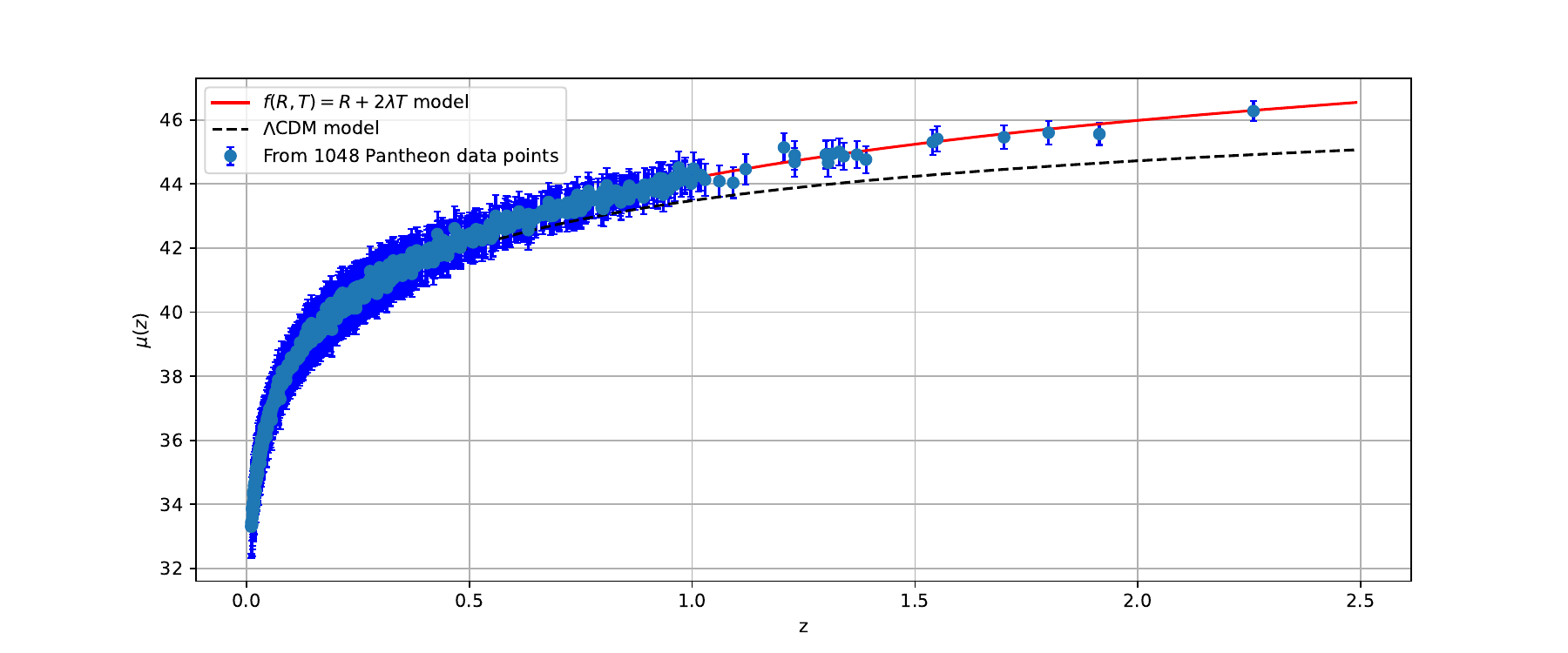}
\caption{Evolution of the distance modulus $\mu(z)$ with redshift $z$ in comparison to $\Lambda$CDM model.}
\label{Mu}
\end{figure}	

\begin{figure}[H]
\centering
\includegraphics[scale=0.45]{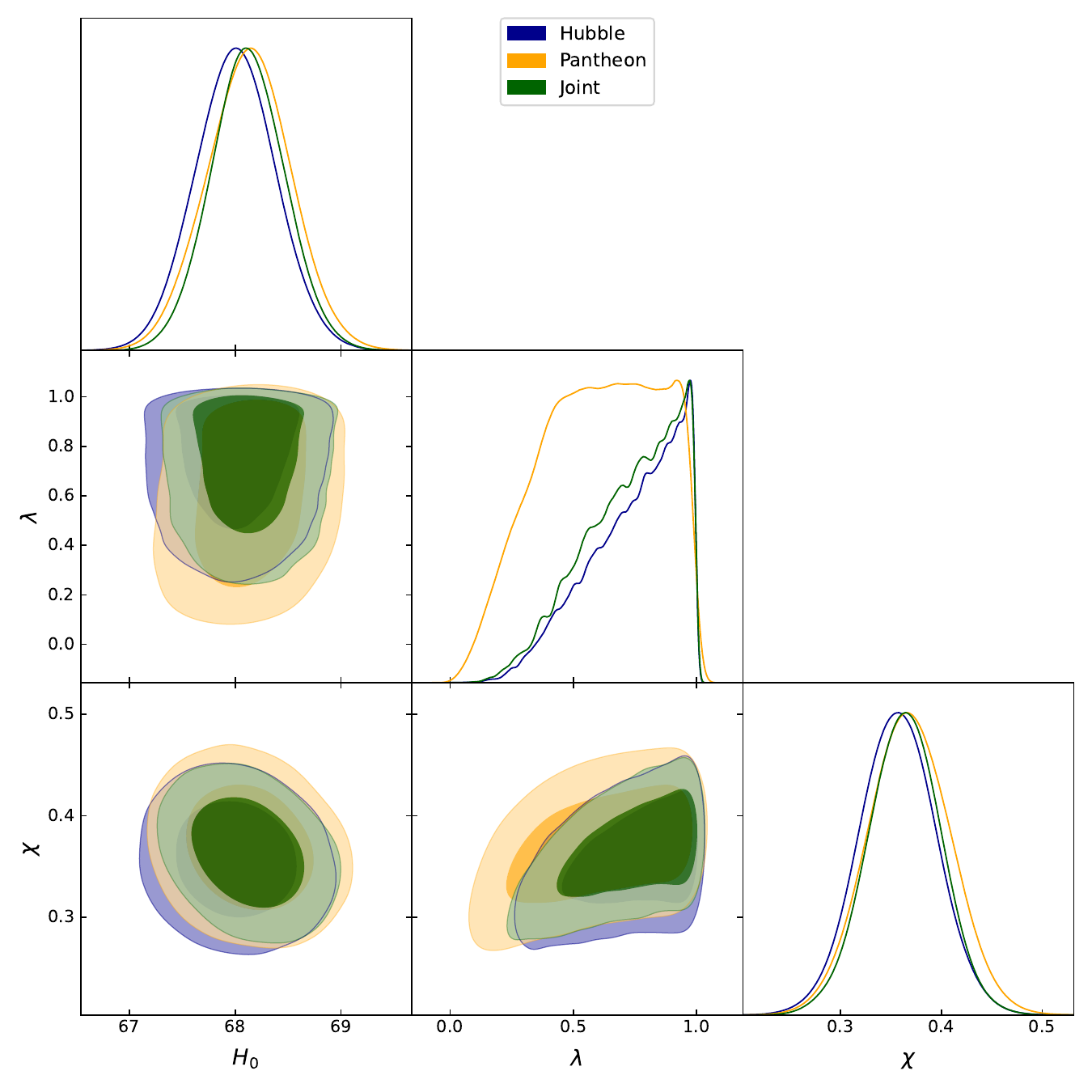}
\caption{Likelihood contours and parameter constraints with $Hubble$ and $Pantheon$ datasets.}
\label{Combine}
\end{figure}

\begin{table*}[!htbp]
\begin{center}
\begin{tabular}{l c c c c c c}
\hline\hline 
$datasets$              & $H_{0}$ ($km/s/Mpc$) & $\lambda$ & $\chi$ & $q_{0}$ & $z_{tr}$ & $\omega_{0}$ \\
\hline

$Hubble$ & $68.01_{-0.73}^{+0.75}$  & $0.74^{+0.26}_{-0.36}$  & $0.357^{+0.075}_{-0.075}$ & $-0.65^{+0.04}_{-0.04}$ & $0.54^{+0.05}_{-0.04}$ &  $-0.65^{+0.05}_{-0.05}$\\

$Pantheon$   & $68.13_{-0.78}^{+0.77}$  & $0.61_{-0.42}^{+0.39}$  & $0.368^{+0.080}_{-0.080}$ & $-0.63^{+0.03}_{-0.03}$ & $0.51^{+0.03}_{-0.03}$& $-0.64^{+0.04}_{-0.04}$\\

$Joint$   & $68.12_{-0.68}^{+0.69}$  & $0.73_{-0.36}^{+0.28}$  & $0.363^{+0.069}_{-0.072}$ & $-0.64^{+0.03}_{-0.03}$ & $0.53^{+0.04}_{-0.03}$ & $-0.64^{+0.04}_{-0.04}$\\

\hline\hline
\end{tabular}
\caption{Cosmological parameter constraints from MCMC analysis: $Hubble$ and $Pantheon$ datasets.}
\label{tab}
\end{center}
\end{table*}

\end{widetext}

\section{Analysis of Cosmological Parameters from Joint Observational Data}
\label{sec5}

In this study, our analysis primarily focused on the joint dataset consisting of both the $Hubble$ and $Pantheon$ data. This decision was based on the observation that other datasets exhibit similar behavior and trends as the joint solution. The evolution of various cosmological parameters, including the density parameter, pressure, deceleration parameter, and effective EoS parameter, is investigated based on the joint observational data.

\begin{figure}[H]
\includegraphics[scale=0.7]{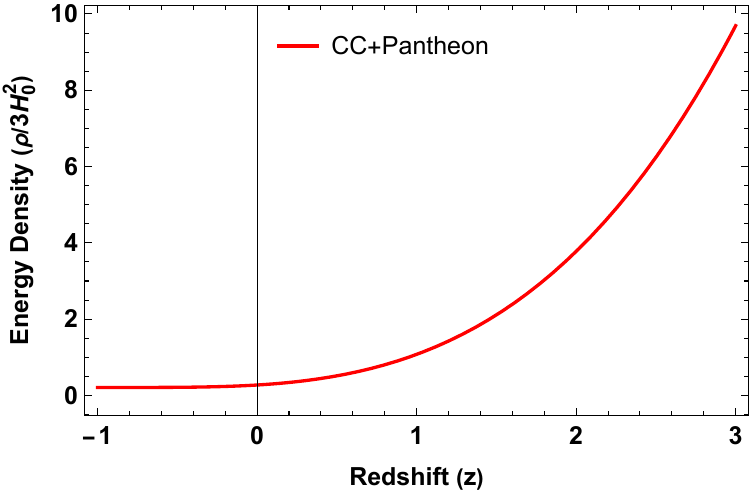}
\caption{Evolution of the density parameter $\rho /3H_{0}^2$ with redshift $z$}
\label{F_rho}
\end{figure}

\begin{figure}[H]
\includegraphics[scale=0.72]{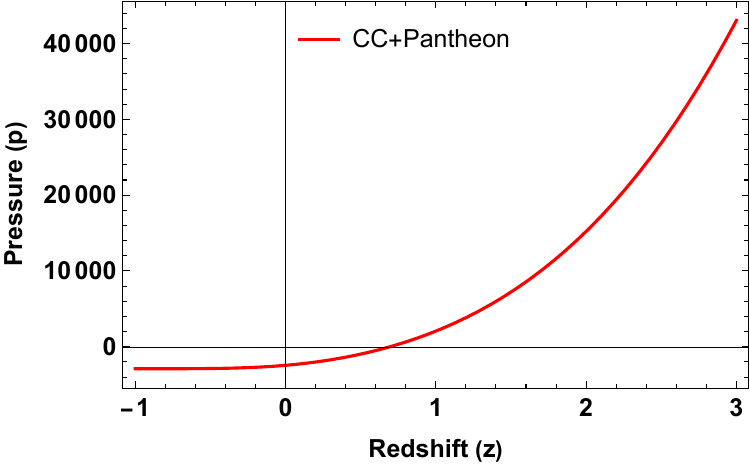}
\caption{Evolution of the pressure $p$ with redshift $z$}
\label{F_p}
\end{figure}

From Fig. \ref{F_rho}, it is clear that the density parameter maintains a positive value throughout the evolution of the Universe, while increasing as the redshift $z$, increases. Initially, at high redshifts, the density parameter has a significant positive value, gradually approaching zero as $z\to-1$. This behavior aligns with our expectations and confirms the overall consistency of our model. On the other hand, the pressure in Fig. \ref{F_p} exhibits an interesting behavior. It starts from a large positive value at high redshifts and progressively transitions to negative values in the present epoch. This shift towards negative pressure is in line with the presence of DE, which is responsible for driving the accelerated expansion of the Universe. The observed negative pressure, consistent with the notion of DE, provides empirical support for the accelerated expansion and reinforces the validity of our model.

\begin{figure}[H]
\includegraphics[scale=0.7]{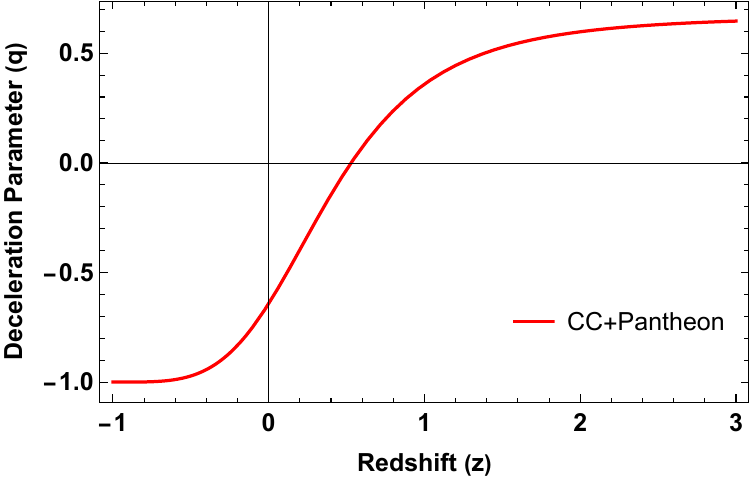}
\caption{Evolution of the deceleration parameter $q$ with redshift $z$}
\label{F_q}
\end{figure}

The deceleration parameter serves as a measure of the Universe's evolution. It characterizes the transition from an early decelerating phase ($q > 0$) to the current accelerating phase ($q < 0$) in cosmological models. The classification of models is based on the time dependence of the deceleration parameter. Observations in recent times have provided strong evidence that the present Universe is indeed experiencing an accelerated phase of expansion. The present value of the deceleration parameter is $q_0=-0.64^{+0.03}_{-0.03}$ \cite{Almada}, which falls within the range of $-1 \leq q < 0$, indicating the transition from deceleration to acceleration. This finding aligns with the established understanding that the expansion of the Universe is currently accelerating, contrary to the earlier decelerating phase (see Fig. \ref{F_q}). The transition from deceleration to acceleration is characterized by a specific transition redshift $z_{tr}$. In our study, we have analyzed the joint dataset of the $Hubble$ and $Pantheon$ data and determined the transition redshift to be $z_{tr}=0.53^{+0.04}_{-0.03}$. The derived value of the transition redshift in our model aligns well with the observational data \cite{Capozziello2014, Capozziello2015, Farooq2017}.

\begin{figure}[H]
\includegraphics[scale=0.7]{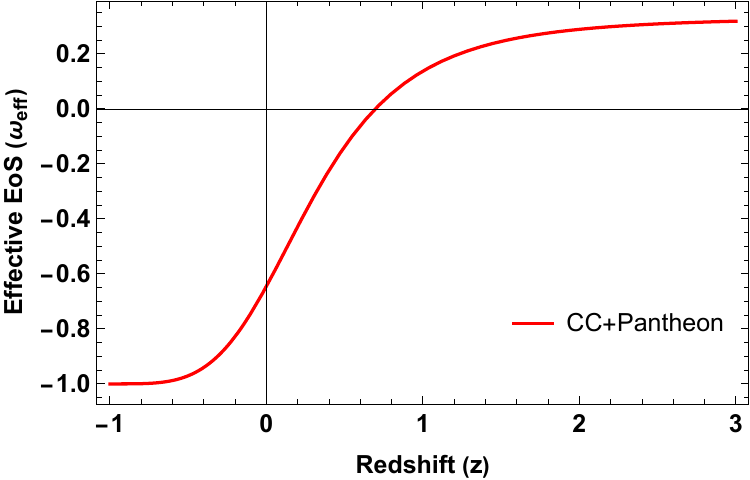}
\caption{Evolution of the effective EoS parameter $\omega_{eff}$ with redshift $z$}
\label{F_EoS}
\end{figure}

As mentioned earlier, the EoS parameter is a useful tool for distinguishing between different epochs of accelerated and decelerated expansion in the Universe. It provides insights into the nature of the cosmic constituents. By examining the value of $\omega$, we can identify various phases:
\begin{itemize}
    \item When $\omega = 1$, it corresponds to a stiff fluid, characterized by high energy density and pressure. This phase is associated with rapid expansion.
    \item For $\omega = \frac{1}{3}$, we have the radiation-dominated phase. During this period, radiation is the dominant component in the Universe, with its energy density decreasing more slowly compared to the expansion.
    \item The matter-dominated phase is characterized by $\omega = 0$. In this epoch, matter (both baryonic and dark matter) becomes the dominant component, contributing significantly to the energy density. The expansion rate is slower than in the previous phases.
    \item During the acceleration phase, the EoS parameter takes on values less than $-1/3$, indicating the dominance of a component with negative pressure, commonly referred to as DE. In addition, the accelerating phase of the Universe can be described by three possible states: the cosmological constant ($\omega =-1$), quintessence ($-1<\omega<-1/3$), and the phantom era ($\omega <-1$). 
\end{itemize}

In addition to the previous discussion, Fig. \ref{F_EoS} clearly illustrates that the effective EoS parameter $\omega_{eff}$ for the model parameters constrained by the joint dataset is less than $-1/3$, indicating a quintessence DE component and implying an accelerating phase of the Universe. Furthermore, it is worth noting that in our model, the effective EoS parameter does not cross the phantom divide at $\omega =-1$.  This finding has important implications as crossing the phantom divide would have significant consequences for the future evolution of the Universe, potentially leading to instabilities and a "Big Rip" scenario \cite{Briscese2007}. The fact that our model remains in the quintessence region provides some theoretical stability and avoids these extreme outcomes. 

Through the process of fitting our model to the observational data, we have obtained the present value of the effective EoS parameter as $\omega_{eff}=-0.64^{+0.04}_{-0.04}$ for the joint dataset \cite{Gruber}. This value represents the best-fit estimation based on the constrained values of the model parameters and provides valuable information about the nature of the cosmic expansion at the present epoch.

\section{Stability Analysis: Sound Speed}
\label{sec6}

The stability of a DE model can be evaluated through the analysis of the square of the sound speed $v_{s}^2$. Specifically,  a positive value of $v_{s}^2$ signifies the stability of the model, while a negative value implies a state of instability \cite{Peebles}. The expression for $v_{s}^2$ is given by
\begin{equation}
    v_{s}^2=\frac{ \partial p}{\partial \rho}.
\end{equation}

For the model under consideration, the expression for the square of the sound speed is given by
\begin{equation}
    v_{s}^2(z)=\frac{6 \lambda +(1+2 \lambda) \chi  (1+z)^4+1}{10 \lambda +3 (1+2 \lambda) \chi  (1+z)^4+3}.
\end{equation}

\begin{figure}[H]
\includegraphics[scale=0.7]{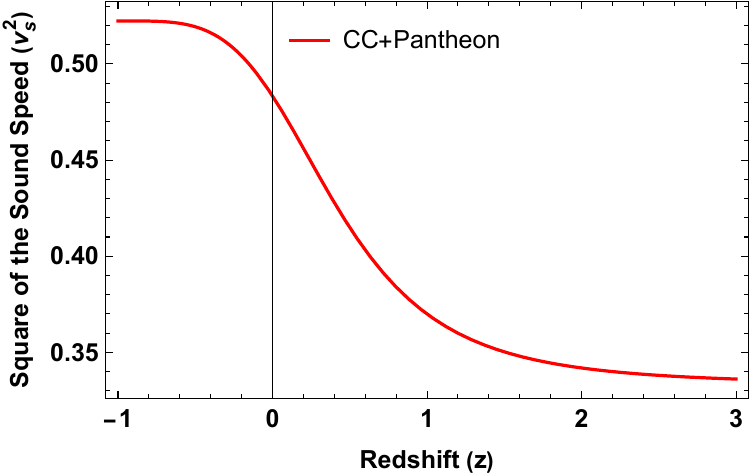}
\caption{Evolution of the square of the sound speed $v_{s}^2$ with redshift $z$}
\label{F_vs}
\end{figure}

Fig. \ref{F_vs} illustrates the evolution of the squared sound speed ($v_{s}^2$) for the model, as depicted by the joint dataset. Notably, the plot reveals that $v_{s}^2$ maintains a consistently positive value across all redshift values. This noteworthy observation indicates the stability of the cosmological model under scrutiny \cite{Huang2019}.

\section{Conclusion}
\label{sec7}

As the exploration of new gravity theories continues to expand, it becomes crucial to subject them to rigorous testing in order to assess their viability in describing the elusive dark sector of the Universe. One such promising theory is $f(R, T)$ gravity, which combines the curvature scalar $R$ with the trace of the energy-momentum tensor $T$. This novel approach offers a unique perspective on gravity and holds the potential to provide deeper insights into the nature of DE and its role in shaping the evolution of the cosmos. 

To initiate our analysis, we adopted a specific functional form for $f(R, T)$, namely $f(R, T) = R + 2\lambda T$, where $\lambda$ represents a free parameter. In contrast to preceding research endeavors, our study distinguishes itself through several pivotal facets, culminating in unique findings and insights. While related investigations have delved into the realm of late-time cosmology within the context of $f(R, T)$ gravity (as noted in Refs. \cite{Pradhan1,Pradhan2,Pradhan3,Pradhan4}), our work advances distinctively by adopting specific functional forms for the EoS parameter. In particular, for the purpose of solving the field equations governing the Hubble parameter $H$, we have employed a carefully chosen parametric form for the effective EoS parameter as a function of redshift $z$: $\omega_{eff}=\frac{1}{3} \left[1-\frac{4}{1+\chi  (1+z)^4}\right]$, where $\chi$ is a constant. This parametric form has advantages for depicting the Universe's evolution across epochs. It follows the standard cosmological model, with specific values at different redshifts. At present, the EoS parameter is expected to be less than $-1/3$, indicating dark energy domination. In the past, $\omega_{eff}$ approaches zero, consistent with matter domination. At larger redshifts, $\omega_{eff}$ converges to $1/3$, representing radiation domination. Furthermore, we compared the Hubble parameter $H(z)$ with the most recent observational datasets, specifically the $Hubble$ and $Pantheon$ datasets, to constrain the model parameters $H_0$, $\lambda$, and $\chi$. The results, including the best-fit ranges of the model parameters along with the $1-\sigma$ and $2-\sigma$ likelihood contours, are presented in Fig. \ref{Combine} and summarized in Tab. \ref{tab}. To investigate the evolution of cosmological parameters  such as the density parameter, pressure, deceleration parameter, and effective EoS parameter, our analysis focused on the joint dataset of the $Hubble$ and $Pantheon$ data.

Figs. \ref{F_rho} and \ref{F_p} demonstrate that the density parameter decreases as the Universe expands, while the pressure exhibits a negative behavior. Also, our analysis reveals a smooth transition of the deceleration parameter $q$ from a decelerated to an accelerated period of expansion (see Fig. \ref{F_q}). The effective EoS parameter shows a negative behavior (see Fig. \ref{F_EoS}), indicating an accelerating Universe and quintessence of DE. The present values of the deceleration parameter, $q_0=-0.64^{+0.03}_{-0.03}$, and the effective EoS parameter, $\omega_{eff}=-0.64^{+0.04}_{-0.04}$, obtained in our analysis are in better agreement with the most recent cosmological observations \cite{Almada,Capozziello2014, Capozziello2015,Farooq2017,Gruber}. It is noteworthy that the behavior of the deceleration parameter follows a similar pattern to that of the effective EoS parameter. This observation suggests a close connection between the dynamics of cosmic acceleration and the underlying nature of DE. The consistent behavior of these two parameters provides further support for the validity and reliability of our analysis.

Also, we have conducted a thorough investigation into the stability of the model. Stability analysis is crucial to ensure the viability and robustness of the proposed framework. Our findings indicate that the model exhibits stability (see Fig. \ref{F_vs}), indicating its suitability for describing the evolution of the Universe. To conclude, although the $f(R, T) = R + 2\lambda T$ model has drawn inspiration from numerous studies, our present inquiry uniquely revolves around constraining this model using contemporary datasets, culminating in the determination of best-fit values for the coupling parameter $\lambda$. This endeavor serves as a catalyst, sparking and fostering subsequent investigations into this model and its broader integration across diverse realms of cosmology and astronomy. Our study decisively substantiates that the proposed methodology reliably predicts the Universe's accelerated expansion, thereby proffering an alternative avenue to DE. In doing so, it lays a foundational framework by furnishing essential cosmological parameter values for intricate explorations into the profound implications this model may hold.

\section*{Acknowledgments}
This work was supported and funded by the Deanship of Scientific 
Research at Imam Mohammad Ibn Saud Islamic University (IMSIU) 
(grant number IMSIU-RG23008).

\textbf{Data availability} This article does not include any new associated data.


\begin{thebibliography}{999}
\bibitem{Perlmutter/1999} S. Perlmutter et al.,  \textit{Astrophys. J.}, \textbf{517} 377 (1999).

\bibitem{Riess/1998} A.G. Riess et al., \textit{Astron. J.}, \textbf{116} 1009 (1998).

\bibitem{Riess/2004} A.G. Riess et al., \textit{Astophys. J.}, \textbf{607} 665-687 (2004).

\bibitem{Komatsu/2011} E. Komatsu et al., \textit{Astrophys. J. Suppl.}, \textbf{192}, 18 (2011).

\bibitem{Huang/2006} Z.Y. Huang et al., \textit{JCAP}, \textbf{0605}, 013 (2006).

\bibitem{Koivisto/2006} T. Koivisto, D.F. Mota, \textit{Phys. Rev. D}, \textbf{73}, 083502 (2006).

\bibitem{Daniel/2008} S.F. Daniel, \textit{Phys. Rev. D}, \textbf{77} (2008) 103513.

\bibitem{Peebles/1988} P. J. E. Peebles and B. Ratra, \textit{Astrophys. J. Lett.}, \textbf{325}, L17 (1988). 
 
\bibitem{Ratra/1988}  B. Ratra and P. J. E. Peebles, \textit{Phys. Rev. D}, \textbf{37}, 3406 (1988). 
 
\bibitem{Steinhardt/1999}  P. J. Steinhardt, L. Wang, and I. Zlatev, \textit{Phys. Rev. D}, \textbf{59}, 123504 (1999).

\bibitem{Peebles/2003} P. J. E. Peebles and B. Ratra, \textit{Rev. Mod. Phys.}, \textbf{75}, 559 (2003).

\bibitem{Sahni/2000} V. Sahni, A. Starobinsky, \textit{Int. J. Mod. Phys. D}, \textbf{9}, 373 (2000).

\bibitem{carroll04} S. M. Carroll, V. Duvvuri, M. Trodden, M. S. Turner, \textit{Phys. Rev. D}, \textbf{70} 043528 (2004).

\bibitem{nojiri07} S. Nojiri, S. D. Odintsov, \textit{Int. J. Geom. Methods Mod. Phys.}, \textbf{04} 115 (2007).

\bibitem{bertolami07} O. Bertolami, et al., \textit{Phys. Rev. D}, \textbf{75} 104016 (2007).

\bibitem{Dixit} A. Dixit et al., \textit{Ind. J. Phys.}, 1-11 (2023).

\bibitem{bengocheu09} G. R. Bengochea, R. Ferraro, \textit{Phys. Rev. D}, \textbf{79} 124019 (2009).

\bibitem{linder10} E. V. Linder, \textit{Phys. Rev. D}, \textbf{81} 127301 (2010).

\bibitem{Chen} S.H. Chen et al., \textit{Phys. Rev. D}, \textbf{83} 023508 (2011).

\bibitem{Anagnostopoulos} F.K. Anagnostopoulos, S. Basilakos, and E.N. Saridakis, \textit{Phys. Rev. D}, \textbf{100} 083517 (2019).

\bibitem{Jimenez1} J. B. Jiménez, L. Heisenberg, and T. Koivisto, \textit{Phys. Rev. D}, \textbf{98} 044048 (2018).

\bibitem{Jimenez2} J. B. Jiménez et al., \textit{Phys. Rev. D}, \textbf{101} 103507 (2020).

\bibitem{Banerjee} A. Banerjee et al., \textit{Eur. Phys. J. C}, \textbf{81} 1-7 (2021).

\bibitem{Pradhan6} A. Pradhan, D.C. Maurya, A. Dixit, \textit{Int. J. Geom. Methods Mod. Phys.}, \textbf{18} 2150124 (2021).

\bibitem{Harko2011} T. Harko, F. S. N. Lobo, S. Nojiri, and S. D. Odintsov, \textit{Phys. Rev. D}, \textbf{84}, 024020 (2011).

\bibitem{Sahu17} S. K. Sahu, S. K. Tripathy, P. K. Sahoo, A. Nath, \textit{Chin. J. Phys.}, \textbf{55} 862 (2017).

\bibitem{Mishra16} B. Mishra, S. Tarai, S. K. Tripathy, \textit{Adv. High. Energy. Phys.}, \textbf{2016} 8543560 (2016).

\bibitem{Moraes17} P. H. R. S. Moraes, P. K. Sahoo, \textit{Eur. Phys. J . C}, \textbf{77} 480 (2017).

\bibitem{Yousaf17} Z. Yousaf, M. Ilyas, M. Z. Bhatti, \textit{Eur. Phys. J. Plus}, \textbf{132} 268 (2017).

\bibitem{Pradhan5} A. Pradhan, A. Dixit, and G. Varshney, \textit{Int. J. Mod. Phys. A}, \textbf{37} 2250121 (2022).

\bibitem{Pretel} J.M.Z. Pretel et al., \textit{Chin. Phys. C}, \textbf{46} 115103 (2022).

\bibitem{Silva} F. M. Silva et al., \textit{Eur. Phys. J. C}, \textbf{83} 295 (2023).

\bibitem{Vinutha} T. Vinutha, V. V. Kuncham, S. K. Kolli, \textit{Gen. Relativ. Gravit.}, \textbf{55} 64 (2023).

\bibitem{Bishi} B. K. Bishi, P. V. Lepse, and A. Beesham, \textit{Chin. J. Phys.}, \textbf{81} 162-170 (2023).

\bibitem{Carroll/2003} Sean M. Carroll, M. Hoffman, M. Trodden, \textit{Phys. Rev. D}, \textbf{68}, 023509 (2003).

\bibitem{Gong/2007} Y. Gong, A. Wang, \textit{Phys. Rev. D}, \textbf{75}, 043520 (2007).

\bibitem{Huang/2008} Q-Guo Huang, \textit{Phys. Rev. D}, \textbf{77}, 103518 (2008).

\bibitem{CPL1} M. Chevallier and D. Polarski, \textit{Int. J. Mod. Phys. D}, \textbf{10}, 213 (2001).

\bibitem{CPL2} E. V. Linder, \textit{Phys. Rev. Lett.}, \textbf{90}, 091301 (2003).

\bibitem{JBP} H. K. Jassal, J. S. Bagla, T. Padmanabhan, \textit{Mon. Not. R. Astron. Soc. Lett.}, \textbf{356}, L11-L16 (2005).

\bibitem{MZ} J.-Z. Ma and X. Zhang, \textit{Phys. Lett. B}, \textbf{669}, 233 (2011).

\bibitem{Lapola} M. M. Lapola et al., \textit{Chin. J. Phys.}, \textbf{72}, 159-175 (2021).

\bibitem{Tanabashi} M. Tanabashi et al., \textit{Phys. Rev. D}, \textbf{98}, 1-1898 (2018).

\bibitem{Mussatayeva} A. Mussatayeva, N. Myrzakulov, and M. Koussour, \textit{Phys. Dark Universe}, \textbf{42}, 101276 (2023).

\bibitem{Mukherjee} A. Mukherjee, \textit{ Mon. Notices Royal Astron. Soc.}, \textbf{460}, 273-282 (2016).

\bibitem{Koussour1} M. Koussour and A. De, \textit{Eur. Phys. J. C}, \textbf{83}, 400 (2023).

\bibitem{Arora} S. Arora, A. Parida and P.K. Sahoo, \textit{Eur. Phys. J. C}, \textbf{81}, 1-7 (2021).

\bibitem{Moresco/2015} M. Moresco, \textit{Mon. Not. R. Astron. Soc.}, \textbf{450}, L16 (2015).

\bibitem{Moresco/2018} M. Moresco et al., \textit{Astrophys. J.}, \textbf{868}, 84 (2018).

\bibitem{Scolnic/2018} D. M. Scolnic et al., \textit{Astrophys. J.}, \textbf{859}, 101 (2018). 

\bibitem{Chang/2019} Z. Chang et al., \textit{Chin. Phys. C}, \textbf{43}, 125102
(2019).

\bibitem{Mackey/2013} D. F. Mackey et al., \textit{Publ. Astron. Soc. Pac.}, \textbf{125}, 306 (2013).

\bibitem{Planck2020} Planck Collaboration, \textit{Astron. Astrophys.}, 
\textbf{641}, A6 (2020).

\bibitem{Almada} S. A. Hernandez-Almada, et al, \textit{Eur. Phys. J. C}, 
\textbf{79}, 12 (2019).

\bibitem{Capozziello2014} S. Capozziello, et al., \textit{Phys. Rev. D}, 
\textbf{90}, 044016 (2014).

\bibitem{Capozziello2015} S. Capozziello, O. Luongo, and E. N. Saridakis, \textit{Phys. Rev. D}, 
\textbf{91}, 124037 (2015).

\bibitem{Farooq2017} O. Farooq, et al., \textit{ Astrophys. J.}, \textbf{835}, 26 (2017).

\bibitem{Briscese2007} F. Briscese, et al., \textit{Phys. Lett. B}, \textbf{646}, 105-111 (2007).

\bibitem{Gruber} C. Gruber, and O. Luongo, \textit{Phys. Rev. D} \textbf{89}, 103506 (2014).

\bibitem{Peebles} P.J.E. Peebles, and B. Ratra, \textit{Rev. Mod. Phys.}, \textbf{75}, 559 (2003).

\bibitem{Huang2019} Q. Huang et al., \textit{Class. Quantum Gravity}, \textbf{36}, 175001 (2019).

\bibitem{Pradhan1} A. Pradhan, G. Goswami, and A. Beesham, \textit{J. High Energy Phys.}, \textbf{38}, 12-21 (2023).

\bibitem{Pradhan2} A. Pradhan, G. Goswami, and A. Beesham, \textit{Int. J. Geom. Methods Mod. Phys.}, \textbf{20}, 2350169 (2023).

\bibitem{Pradhan3} A. Pradhan, G. Goswami, and S. Krishnannair, \textit{Eur. Phys. J. Plus
}, \textbf{138}, 1-12 (2023).

\bibitem{Pradhan4} A. Pradhan et al., \textit{Astron. Comput.}, \textbf{44}, 100737 (2023).





\end{thebibliography}
\end{document}